# Mechanical analogies for gravitation

Valery P. Dmitriyev

*Lomonosov University, Moscow, Russia*
*P.O.Box 160, Moscow 117574, Russia*
*e-mail: dmitr@cc.nifhi.ac.ru*

**Abstract**

A quasielastic model of gravitation is developed. Gravitational field is modeled by the flow of the transient point dilatation, which is supposed to be continually emitted by a discontinuity of the turbulent fluid. Gravitational attraction arises from the contact interaction of the dilatation centers. Gravitational wave is viewed microscopically as the axisymmetric wave propagating at a high speed along a vortex tube.

PACS 04.50.+h – Alternative theories of gravity.
PACS 03.40.Dz – Theory of elasticity.
PACS 47.32.Cc – Vortex dynamics.

## 1  Introduction

Mechanical analogies of physical fields and particles can be found among solutions to Euler equation. Undoubtedly, it is the Bjerknes force acting between the two pulsating bubbles immersed in a fluid [1] that should be taken as a microscopic mechanical model of gravitating particles. However, as before, in the current report we will abide by the mesoscopic approach. For, it is directly comparable with the conventional theories. By its structure, the mesoscopic models belong to elasticity.

The whole scheme looks as follows. A turbulent ideal fluid serves as a substratum for physics. In the ground state, the turbulence is supposed to be homogeneous and isotropic. Disturbances of the turbulence model physical fields [2]. Particles originate themselves from the voids in the fluid [3]. The stationary discontinuities formed from the voids generate in the fluid gradient pressure fields and thus affect each other distantly. The pressure fields vary with the distance as $1/r$. These provide us with the model of the electrostatic interaction [3].

The mechanism of gravitation is based on the contact interaction of the dilatation centers [5]. The latter is just what the present report is devoted to.

## 2 Interactions

We consider defects of an elastic medium. There is only one way for two defects affecting each other statically. This is a source of the displacement field $\mathbf{s}(\mathbf{x},t)$ interacting with a source of the pressure field $p(\mathbf{x},t)$. The work done against the perturbation pressure field $dp$ while producing the displacement field $\mathbf{s}$ is given by[*]

$$E_{12} = \int \mathbf{s} \cdot \nabla dp \, d^3x \\ = -\int dp \, \nabla \cdot \mathbf{s} \, d^3x . \quad (2.1)$$

Let the defect be a center of dilatation:
$$\nabla \cdot \mathbf{s} = \Delta V \, \delta(\mathbf{x} - \mathbf{x}'), \quad (2.2)$$

where $\Delta V = \text{const}$ and $\mathbf{x}'$ is the center's location. From the linear stress-strain relation we find the pressure $dp'$ needed in order to create the dilatation center (2.2)

$$dp' = Vc_g^2 \nabla \cdot \mathbf{s} = Vc_g^2 \Delta V \, \delta(\mathbf{x} - \mathbf{x}'), \quad (2.3)$$

where $V$ is the medium density and $c_g$ the speed of the longitudinal wave in the medium. So, two dilatation centers do not interact with each other distantly.

However, there is a special case of the immediate contact.

## 3 Dispersion of a singularity

A dilatation center is supposed to be split and dispersed in the stochastic medium [3-5]. This is described by that we take the strength $\Delta V$ of the inclusion as a function of $\mathbf{x}'$ and $t$. That gives on the place of (2.2):

$$\nabla \cdot \mathbf{s} = c(\mathbf{x}',t), \quad (3.1)$$

where
$$\int c(\mathbf{x}',t) d^3x = \Delta V = \text{const} .$$

Accordingly, we have instead of (2.3)
$$dp' = Vc_g^2 c(\mathbf{x}',t) . \quad (3.2)$$

Substituting (3.1), (3.2) to (2.1) gives the energy of interaction of the two clouds of the point dilatation

$$E_{12} = -Vc_g^2 \int c_1(\mathbf{x}',t) c_2(\mathbf{x}',t) d^3x' . \quad (3.3)$$

It should be stressed here that the sign in (3.3) refers to interaction of the two "external forces" (2.3) applied to support the centers of dilatation.

---

[*] In the present report symbol $p$ refers to pressure, while in [2,3] it was used for specific pressure.
Besides, though nowhere defined explicitly, all the quantities pertain to the turbulence averages.

## 4 Gravitation

A particle is modeled by a point discontinuity of the turbulent substratum. We see it as a permanent source of the flow of the point dilatation. Supposing that the intensity $G$ of the source is extremely small, the distribution of the point dilatation in the stationary flow can be approximated [5] by

$$c(\mathbf{x}') = \frac{G}{4\pi c_g} \frac{1}{(\mathbf{x}' - \mathbf{x}_0)^2}, \qquad (4.1)$$

where $\mathbf{x}_0$ is the source's location. Here the velocity of the flow has been taken as the speed $c_g$ of the longitudinal wave in the medium. Substituting (4.1) in (3.3) we get

$$E_{12} = -Vc_g^2 \frac{G_1 G_2}{(4\pi c_g)^2} \int \frac{d^3 x'}{(\mathbf{x}' - \mathbf{x}_1)^2 (\mathbf{x}' - \mathbf{x}_2)^2} = -\frac{\rho V}{16} \frac{G_1 G_2}{|\mathbf{x}_1 - \mathbf{x}_2|}.$$

The Newton's gravitational law is obtained from the latter if we postulate

$$G = gV,$$

where $V$ is the volume of the particle's core and $g = \text{const}$. Then, defining the particle's mass by

$$m = W,$$

we get finally

$$E_{12} = -\frac{\rho}{V} \left(\frac{g}{4}\right)^2 \frac{m_1 m_2}{|\mathbf{x}_1 - \mathbf{x}_2|}.$$

## 5 Gravitation and electrostatics

We see the substratum as a dispersion of voids in a turbulent fluid. Then, there are two kinds of compressibility that should be distinguished.

The macroscopic compressibility

$$d_r p = c_r^2 d_r V \qquad (5.1)$$

is concerned with a redistribution of the empty space in the medium. This kind of elasticity is responsible for electromagnetism, the Lorentz gauge being accounted for in (5.1) [4]. Really, $d_r p$ is determined by hydrodynamics and should be calculated from Euler equation. Being equilibrated by the Reynolds stresses, those disturbances of the pressure can exist as a steady formation thereby giving rise to electrostatics [3].

The microscopic compressibility is due to variation in the fine-scale density:

$$d_g p = c_g^2 d_g V. \qquad (5.2)$$

In the medium, which is microscopically incompressible though discontinuous, this kind of elasticity occurs only in a singular mode:

$$d_g V = -V \Delta V \, \delta(\mathbf{x}-\mathbf{x}'),$$

that is implied by (2.3) with (5.2). The external force (2.3) is needed in order to support the dilatation center. Therefore, the microscopic elasticity can be developed only in dynamics e.g. in a wave. In the mesoscopic view, this means that the dilatation center ought to be in a motion.

The speed of the longitudinal wave in the substratum is much greater than the speed of the transverse wave

$$c_g \gg c_r. \qquad (5.3)$$

Ideally, we have

$$c_g \to \infty.$$

## 6  Comparison with general relativity

In its mathematical structure general relativity is known [6] to be a subalgeabra of elasticity. In this event, the space metric from the geometric model corresponds to the nonsymmetrical strain tensor from the substratum model. In the relativistic theory of gravitation, the general criterion that physical fields are present is that the terms of the curvature tensor are nonvanishing. The presence of discontinuities in an elastic medium is indicated by that the terms of Kroner's incompatibility tensor are different from zero. The latter mathematical construction corresponds exactly to Ricci tensor. However, at this point the similarity between the elastic model of gravitation and general relativity lets out. In general relativity the element $g_{00}-1$ of metric tensor is taken as an immediate measure of Newton gravitational potential. In terms of the linear-elastic substratum this element corresponds to the potential component $\nabla \cdot \mathbf{s}$ of the strain tensor and is described by the volume density (3.1) of the dilatation centers. As you see, in the mechanical model gravitational potential is measured by (3.3), which is the convolution of those densities.

## 7  Microscopic support

Microscopically, the substratum can be viewed as the vortex sponge. It is represented by a heap of randomly oriented rectilinear vortex tubes [7]. There are two kinds of the wave propagating along a vortex tube: the torsional (helical or kink) wave and the axisymmetic (area varying) wave [8].

The torsional wave is concerned with the deviation of the vortex filament from the rectilinear form. It is taken as a model of the electromagnetic wave. On the other side, the disturbance of the rectilinear configuration associated with a redundant segment of the vortex tube intruded models a particle. For instance, it may be a loop on the vortex filament [9]. The structural similarity indicated is just what the wave-particle dualism comes from. The dynamical soliton on the vortex filament generates in the fluid the far-field sound pressure, which stretches away as $1/r$ [10]. That may be taken as a mechanical model of the electrostatic field produced in "vacuum" by an electric charge [2,3].

The axisymmetric wave represents a different type of the distortion. This is the wave of expansion of the tube's cross section or simply a moving swelling of the vortex tube that we are interested in. It is taken as a model of the gravitational wave. Experimentally [8], the speed of the axisymmetric wave exceeds enormously the speed of the torsional wave. Ideally, i.e. in an incompressible medium, it propagates instantaneously. That is from we have the inequality (5.3). This kind of the wave is responsible for the "wave-function collapse" and for instantaneous transmission of the quantum correlations.

The axisymmetric wave was modeled above mesoscopically by the flow of the transient point dilatation. The "external force" (2.3), which is applied to the medium in order to support the nonstationary dilatation centers, just corresponds to the inertial force in the axisymmetric wave.